\documentclass[pre,twocolumn,superscriptaddress,amsmath,amssymb]{revtex4}
\usepackage{graphicx}
\usepackage{dcolumn}
\usepackage{bm}
\usepackage{color}
\usepackage{natbib}

\begin{document}

\title{Growth saturation of unstable thin films on transverse-striped hydrophilic-hydrophobic micropatterns}

\author{\firstname{R.} \surname{Ledesma-Aguilar}}%
\affiliation{Departament d'Estructura i Constituents de la
  Mat\`eria. Universitat de Barcelona,
Avinguda Diagonal 647, E-08028 Barcelona, Spain}
\affiliation{The Rudolf Peierls Centre for Theoretical Physics, University of Oxford,
1 Keble Road, Oxford OX1 2HQ, United Kingdom}
\author{\firstname{A.} \surname{Hern\'andez-Machado}}
\affiliation{Departament d'Estructura i Constituents de la
  Mat\`eria. Universitat de Barcelona,
Avinguda Diagonal 647, E-08028 Barcelona, Spain}
\author{\firstname{I.} \surname{Pagonabarraga}}
\affiliation{Departament de F\'isica Fonamental.  Universitat de
  Barcelona, Avinguda Diagonal 647, E-08028 Barcelona, Spain}
\date{\today}

\begin{abstract}
Using three-dimensional numerical simulations, we demonstrate the growth saturation of 
an unstable thin liquid film on micropatterned hydrophilic-hydrophobic substrates.  
We consider different transverse-striped micropatterns, characterized by the total fraction of hydrophilic 
coverage and the width of the hydrophilic stripes. We compare the growth of the film on the micropatterns 
to the steady states observed on homogeneous substrates, which correspond to a saturated sawtooth and growing 
finger configurations for hydrophilic and hydrophobic substrates, respectively.
The proposed micropatterns trigger an alternating fingering-spreading dynamics of the film, which leads to a 
complete suppression of the contact line growth above a critical fraction of hydrophilic stripes.  
Furthermore, we find that increasing the width of the hydrophilic stripes slows down the advancing front, giving 
smaller critical fractions the wider the hydrophilic stripes are.   Using analytical approximations, we quantitatively 
predict the growth rate of the contact line as a function of the covering fraction, and predict the threshold fraction for 
saturation as a function of the stripe width.  
\end{abstract}

\maketitle

\section{Introduction}
\label{sec:TS}
The growth of patterns in forced thin liquid films on solid substrates is an ubiquitous process of both biological 
and technological relevance which has attracted the attention of the scientific community over the last decades 
\cite{Craster-RevModPhys-2009,Bankoff01}.  The manipulation of thin films is important in microfluidics, where the 
increasing miniaturization of fluidic devices demands the development of new and simple ways to handle small volumes 
of liquid.  Due to their smaller 
internal hydrodynamic resistance and increased surface area, thin films can be used 
to overcome problems associated to other geometries, such as microchannels~\cite{Dietrich01}. 
Both experimental and theoretical studies indicate 
the possibility of using clever structuring of microfluidic devices to orient the growth of films 
\cite{Troian04,Diez03,Diez04,Marshall01,Marshall-ComputFluids-2005,Ledesma03} 
or to control the motion of interfaces in confined microfluidic chambers \cite{Mognetti-SoftMatter-2010}. 

Still, a remaining challenge is how to control the intrinsic growth of thin liquid films in the microscale.
When a thin film of liquid is forced to spread on a dry solid substrate, there is a net accumulation of fluid close to 
the contact line, which corresponds to the apparent position where the liquid, the solid, and the surrounding gas meet.  
As a result, the film profile bulges near its leading edge, generating a capillary ridge.  
For films driven by a body force, such as a constant pressure gradient,  variations in the size of the ridge along the 
leading edge of the film have the effect of 
increasing the hydrostatic pressure locally.  This introduces a driving mechanism that amplifies small perturbations to the contact 
line, which are relaxed by the surface tension \cite{Brenner01}.  Such competition of driving and restoring forces gives 
rise to a linear instability \cite{Troian01,Troian02,Troian03}, which, far from a flat advancing front, leads
to the formation of non-linear structures, such as the familiar water rivulets observed in the shower. 

Given the strong interaction with the solid, it is unsurprising that the instability is affected by the wetting properties 
of the liquid \cite{Huppert-Nature-1982a,Silvi01,deBruyn01,deBruyn02}.  Wetting interactions 
determine the equilibrium contact angle of the fluid-fluid interface with the solid.  As a consequence, the 
size and shape of the capillary ridge, which control the driving destabilizing force, depend on wetting.   
For hydrophilic substrates, the contact angle of the interface profile is small, making the capillary ridge 
relatively thin.  The instability is thus weakened \cite{Ledesma03}. Remarkably, if the size of the ridge is 
sufficiently small, the restoring capillary pressure is strong enough to balance the driving force after the early stages of growth. 
The interface thus stops growing, but still retains a curved saturated shape reminiscent of a sawtooth pattern 
\cite{Silvi01}.  In contrast, for hydrophobic substrates the net accumulation of mass at the ridge is large enough 
to outweigh the effect of surface tension, and the emerging film patterns grow steadily, with a shape that resembles 
that of fingers \cite{deBruyn01,deBruyn02}.   Crucially, the lateral lengthscale of the patterns, $\Lambda_{max}$, is 
of the order of the thickness of the thin film.  Therefore, in a microfluidic device one expects that the lengthscale of 
emerging patterns is also of the order of microns.  

Our aim in this paper is to demonstrate the exploitation substrate heterogeneity to control the growth of the contact line, 
up to complete growth saturation. This is appealing to many situations in which fingering 
is undesirable, for example in coating processes and in `chemical channeling' \cite{Dietrich01}, where the desired lateral 
lengthscale of the film needs not to be constrained by the intrinsic lengthscale of the instability.  We shall thus analyze the 
effect of a simple, yet effective, substrate pattern that consists of hydrophilic-hydrophobic stripes oriented transversely to 
the direction of flow, as shown in Fig.~\ref{fig:Geometry}.   As we shall see below, the lengthscale of the micropattern is important, 
and has a significant impact already at scales comparable to the micron-sized films. 
Still, and despite its simplicity, such a pattern has not yet been studied.  
In a previous study \cite{Ledesma03}, we have demonstrated that the growth of the thin film can be tuned
by patterning the solid substrate with transverse or checkerboard arrangements of hydrophilic-hydrophobic domains.   
Previous experimental and numerical studies of flow patterning \cite{Troian04,Diez03,Diez04,Marshall01} have 
mainly focused on the effect longitudinal tracks of varying flow resistance \cite{Troian04} or wetting angle 
\cite{Diez03,Diez04,Marshall01}.  The overall experiments and numerical simulations
evidence is that substrate heterogeneity can be used to handle thin films, through preferential orientation over hydrophilic or 
low-resistance domains in the substrate, or to partially control its growth using checkerboard hydrophilic-hydrophobic 
patterns. As we shall show in the following, the proposed transverse pattern introduces new physical mechanisms, 
particularly the transverse spreading of the film, which lead to the desired saturation of the contact line.  

Our approach to this problem will be to perform Lattice-Boltzmann (LB) simulations \cite{Higuera-EPL-1989} of Navier-Stokes 
hydrodynamics coupled to a phase-field binary fluid model \cite{Swift-PRE-1996} to account for the interface dynamics.
Previous theoretical approaches to model forced thin films on chemically heterogeneous substrates
have mainly consisted of sharp-interface formulations based on the Stokes equations \cite{Marshall01,Marshall-ComputFluids-2005,Diez03,Diez04}.  Most of these
studies use the so-called thin-film equations, which are appropriate to model the fluid dynamics
in the limit of small interface slopes.  Within this framework, a common procedure to regularize
the viscous dissipation singularity is to replace the contact line by a thin precursor film that
extends beyond the leading front.  The effect of the heterogeneity can then be included by choosing
a particular model for the precursor.  For example, an heterogeneous disjoining pressure model that
couples to the precursor film \cite{Marshall01}
has been used to study the motion of the film on
longitudinal-striped and disordered substrates, as well as on ordered and disordered isolated spots \cite{Marshall-ComputFluids-2005}.
Alternatively, a similar model consisting of a spatially-varying thickness of the precursor
film to study the motion of the film over chemical patterns has also been used \cite{Diez03,Diez04}.

Our aim in this work is to study the effect of a sharp contrast between hydrophilic and hydrophobic domains in
the proposed micropatterns.  This corresponds to situations in which the static and dynamic contact angles need not
to be small. Such regime falls beyond the limit of applicability of the thin-film equations, demanding the resolution
of the full three-dimensional dynamics.  This is a complicated problem from
the classical point of view, as it involves the motion of two free boundaries: the film surface and the
contact line. Within sharp-interface formulations, boundary integral methods could be used to address this problem,
as has been done to model the effect of chemical heterogeneities on two-dimensional fluid nanodrops \cite{Moosavi-Langmuir-2008}.
However, the mesoscopic approach that we take here constitutes a reliable three-dimensional Navier-Stokes solver
that allows for both small and large contact angles, necessary to model hydrophilic and hydrophobic
interactions.  Because of the diffuse approximation to describe the fluid interface, the method circumvents
the free-boundary problem, regularizes the viscous dissipation singularity at the contact line, and deals naturally
with merging and pinch-off events, which otherwise need a specific rule in sharp interface approximations. We have carried 
out a validation of our model in a previous study, for the case of driven thin films on homogeneous solid substrates 
\cite{Ledesma01}.

The rest of this work is organized as follows.  In Section~\ref{sec:Model} we introduce the mesoscopic model and the 
Lattice-Boltzmann algorithm to perform the numerical simulations.  
We will complement our numerical results with kinematical calculations for the propagation of the interface.  As we shall see from our 
numerical results, presented
in Section~\ref{sec:Results}, the main effect of the transverse hydrophilic stripes on the forced film is to favor 
its lateral spreading on hydrophilic domains, thus slowing down the leading edge of 
the film.  The same lateral spreading ultimately leads to a speed-up of the trailing edge of 
the film. The net result is that the growth rate of the film decreases with increasing 
fraction of hydrophilic stripes in the substrate, until the film saturates above a critical value of 
this fraction.  
As we detail in the discussion and conclusions of this work, presented in Section~\ref{sec:DC}, we
predict that a relatively small fraction of hydrophilic stripes is sufficient to suppress the 
growth of the contact line in experimental realizations.  Our results thus constitute a promising step towards 
the control of interface growth in microfluidic systems.

\section{Model}

\label{sec:Model}

In this paper we shall perform numerical simulation of Navier-Stokes level hydrodynamics coupled to a phase-field representation 
of the viscous fluid film and surrounding low-viscosity phase.  Such modeling is very useful to address the dynamics of immiscible 
fluids, as it has the advantage of circumventing the free-boundary problem associated to sharp-interface representations, 
as for example in boundary integral methods.  Furthermore, the phase-field model regularizes the 
viscous dissipation singularity at the contact line through a diffusive mechanism acting at small scales (of the order 
of the size of the interface).  One thus does not need to specify additional boundary conditions at the contact line.     
In order to integrate the model equations, we will use a Lattice-Boltzmann algorithm.

\subsection{Hydrodynamics and Phase-Field Model}

\begin{figure}[h!]
\begin{centering}
\includegraphics[width=0.45\textwidth]{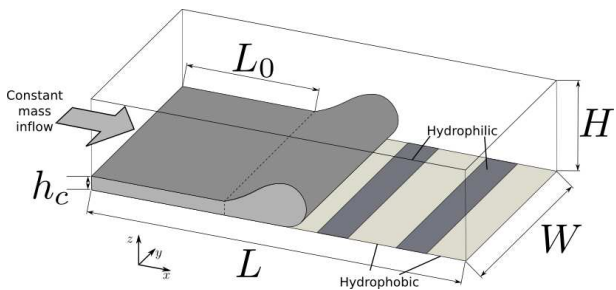}
\end{centering}
\caption{Schematic representation of the system.  A thin film of initial thickness $h$ is forced along the $x$
direction by the action of a constant body force, $\vec f$, on a solid substrate composed of hydrophilic 
and hydrophobic transverse stripes.   The scale of the thin film thickness and of the width of the stripes is 
of the order of microns.  
\label{fig:Geometry}}
\end{figure}

A schematic depiction of the forced thin-film geometry is presented in Fig.~\ref{fig:Geometry}.  We consider two 
immiscible Newtonian fluids of viscosities $\eta_1$ and $\eta_2$ and densities $\rho_1$ and $\rho_2$, respectively. 
The thin film, corresponding to fluid 1, is forced on a solid substrate composed of hydrophilic and hydrophobic 
regions.   In order to distinguish between the two liquids, we introduce a phase field characterized by a concentration 
variable, or order parameter, $\phi$.  The local value of the order parameter thus describes each phase, and varies between 
two volume values across a diffuse interfacial region.   

In equilibrium, the free energy of the system can be written as a functional of the order parameter field, 
$\phi(\vec r)$ and the local density field, $\rho (\vec r)$, 

\begin{equation}
\mathcal F\{\rho,\phi\}=\int_V \mathrm d V  \left( U(\phi,\rho) + 
\frac{\kappa}{2}(\vec\nabla\phi)^2\right) + \int_S \mathrm d S f_S(\phi_S).
\label{eq:FreeEnergy}
\end{equation}

The first integral in Eq.~(\ref{eq:FreeEnergy}) is composed by the volume contributions to the free energy. 
These consist of a double-well potential with an ideal gas term, $U(\phi,\rho) =  A\phi^2/2+B\phi^4/4 + 
(1/3)\rho\ln\rho$, and a square-gradient 
term that penalizes variations of the order parameter, thus allowing for the formation of a diffuse interface.  The 
parameters $A$, $B$ and $\kappa$ control the equilibrium values of the order parameter $\pm \phi_{\mathrm{eq}} = \pm \sqrt{-A/B}$, the 
size of the interface, $\xi = \sqrt{-\kappa /2A}$ and the fluid-fluid surface tension, $\gamma = \sqrt{-8\kappa A^3/9B^2}$.
The second integral in the free energy corresponds to the contribution of fluid-solid interactions.  Here we chose
a Chan model, $f_S(\phi_S)=C\phi_S,$ which assigns the free surface energy $f_S$ according to the local value of the order
parameter at the boundary,  $\phi_S$. The parameter $C$ controls the equilibrium contact angle, $\theta_e$, 
of the fluid-fluid interface in contact with a flat solid wall through the relation
\begin{equation}
\cos \theta_e = \frac{1}{2}\left[-\left(1-C(-\kappa A)^{-\frac{1}{2}}\right)^{\frac{3}{2}}+\left(1+C(-\kappa A)^{-\frac{1}{2}}\right)^{\frac{3}{2}}\right].
\label{eq:ContactAngle}
\end{equation} 

The order parameter and density fields contribute to the chemical potential, $\mu,$ and pressure tensor, $P_{\alpha \beta},$ of the system.  
These are obtained by taking functional derivatives of the free 
energy \cite{pagonabarraga} and have the following expressions, 
\begin{equation}
\mu = A\phi+B\phi^3 - \kappa \nabla^2\phi,
\label{eq:ChemicalPotential}
\end{equation}

\begin{equation}
\begin{array}{rl}
P_{\alpha\beta} & = \left(\frac{1}{3}\rho + \frac{1}{2}A\phi^2+\frac{3}{4}B\phi^4-\kappa\phi\nabla^2\phi-\frac{1}{2}\kappa|\vec \nabla\phi|^2 \right)\delta_{\alpha\beta}\cr
& +\kappa(\partial_\alpha\phi)(\partial_\beta\phi).
\end{array}
\label{eq:PressureTensor}
\end{equation}

Out of equilibrium, the dynamics of the density, $\rho$, velocity, $\vec v$, and order parameter, $\phi$, 
fields are given by the continuity and Navier-Stokes equations, plus a convection-diffusion equation for 
the phase field, {\it i.e.,}
\begin{equation}
\frac{\partial \rho}{\partial t} + \vec \nabla \cdot (\rho \vec v) = 0,
\label{eq:Continuity}
\end{equation}

\begin{equation}
\frac{\partial \vec v}{\partial t} + (\vec v \cdot \vec \nabla) \vec v = -\vec \nabla P - \phi \vec \nabla \mu + 
\eta \nabla^2 \vec v + \vec f ,
\label{eq:NavierStokes}
\end{equation}
and 
\begin{equation}
\frac{\partial \phi}{\partial t} + \vec v \cdot \vec \nabla \phi = M\nabla^2 \mu, 
\label{eq:ConvectionDiffusion}
\end{equation}
respectively.  

The $\phi-$dependent term in the Navier-Stokes equations arises from the concentration dependence 
of the pressure tensor in Eq.~(\ref{eq:PressureTensor}), giving a `chemical' force density, 
$-\phi \vec \nabla \mu$, which plays a similar role as the pressure gradient term in the volume of each phase. 
The remaining terms in the Navier-Stokes equations are the viscous friction term, whose strength is controlled by the local 
viscosity, $\eta$, and to the body force term, $\vec f$.

The evolution of the order parameter is dictated by Eq.~(\ref{eq:ConvectionDiffusion}), which contains 
an advective term caused by the underlying velocity field and a diffusive term, whose strength is controlled 
by the mobility parameter $M$.

\subsection{System Geometry and Boundary Conditions}

In order to ensure the formation of steady patterns, we choose a constant flux configuration for the 
geometry of the system.  We set a rectangular simulation domain of linear dimensions $L\times W\times H$, in the $x$, $y$ and 
$z$ directions, respectively.   The solid-fluid interface, $S$, is parallel to the $x-y$ plane and located at $z=0$ as shown in 
Fig.~\ref{fig:Geometry}.  The thin film is initially at rest and occupies the volume $V_0=L_0\times W\times h_c$.  We fix the 
force term  as $\vec f  = \frac{1}{2}f_x(\phi +1){\hat x}$, so fluid 1 is forced uniformly while fluid 2 is left to evolve 
passively.  This sets up a flow of mean velocity $u = h_c^2 f_x/3\eta_1$ within the film.  At $x=0$, a constant flux in the $x$ direction 
is thus ensured as long as there is a reservoir of 
fluid available to compensate for the downstream motion of the film.  This is achieved by imposing $\partial_x \rho (x=\{0,L\},y,z)=0,$
$\partial_x \vec v (x=\{0,L\},y,z) =\vec 0$ and $\partial_x \phi (x=\{0,L\},y,z)=0.$  Using this approach, we do not find any appreciable variations of the mean velocity of 
the film or the film thickness close to the boundary. 

We fix periodic boundary conditions in the $y$ direction, {\it i.e.}, $\rho(x,y=W,z)=\rho(x,y=0,z)$, $\vec v(x,y=W,z)=\vec v(x,y=0,z)$, 
$\phi(x,y=W,z)=\phi(x,y=0,z)$. At the upper boundary, a shear-free boundary condition is imposed by fixing $\partial_z \vec v (x,y,z=H) =\vec 0$, 
while vanishing density and concentration gradients normal to the boundary are fixed by imposing $\partial_z \rho (x,y,z=H)=0$ and  $\partial_z \phi (x,y,z=H)=0.$

The velocity profile is subject to stick boundary and impenetrability conditions at the solid-fluid surface, {\it i.e.,}
\begin{equation}
\vec v |_S= 0. 
\label{eq:StickBoundaryCondition}
\end{equation}
In macroscopic representations, imposing the stick boundary condition for a moving interface in 
contact with a solid causes a spurious divergence of the viscous stress as one approaches the contact line \cite{Huh01}.  This paradox 
is readily avoided by the phase-field model, through the diffusive term in Eq.~(\ref{eq:ConvectionDiffusion}), which allows for a 
slip velocity that increases as one approaches the contact line \cite{Qian01}, even when Eq.~(\ref{eq:StickBoundaryCondition}) is 
enforced.  Diffusion is driven by variations of the chemical potential profile close to the contact line, over a typical lengthscale 
$l_d$, which scales as $l_d \sim (\eta_1 \xi M/\Delta \phi^2)^{1/4}$ \cite{Yeomans01}.   

An evolution equation for the order parameter should be specified at the solid-fluid 
interface (see \cite{Qian01} for a detailed derivation).  In this paper we shall impose 
the natural boundary condition 
\begin{equation}
\hat n \cdot \vec \nabla \phi|_S = \frac{1}{\kappa}\frac{\mathrm d f_S}{\mathrm d \phi_S}.
\label{eq:NatBCDyn}
\end{equation}
This gives an instantaneous relaxation of the order parameter gradient to its surface value at the wall, 
which corresponds to quasi-equilibrium dynamics for the interface slope at lengthscales comparable to the 
size of the interface \cite{Rolley01}.  By choosing the wetting parameter $C$ in Eq.~(\ref{eq:ContactAngle}), 
we specify the local value of the equilibrium contact angle and hence the desired hydrophilic-hydrophobic pattern.   

The $phi-$dependent term in Eq.~(\ref{eq:NavierStokes}) is most important at the fluid-fluid interface, where it plays the role of 
a Laplace pressure.  This behavior can be verified by integration of Eq.~(\ref{eq:NavierStokes}) over the length of the 
interface (c.f. \cite{bray}) whence one recovers a jump in the normal stress
\begin{equation}
\Delta \mu \sim \frac{\gamma}{R}.
\label{eq:YoungLaplace}
\end{equation}
Eq.~(\ref{eq:YoungLaplace}) corresponds to the Young-Laplace condition with $R$ being the local radius of curvature 
of the interface.  Thus, the concentration dependent term in the Navier-Stokes equation eliminates the need of using 
Eq.~(\ref{eq:YoungLaplace}) explicitly.  Accordingly, the continuity of the tangential stress is recovered in the sharp interface limit.

In order to present our results in a natural scale, we choose units according to the classical thin-film theory 
in the lubrication limit \cite{Craster-RevModPhys-2009,Bankoff01}.   For microscales inertial effects are negligible, and 
the relevant control parameter is the capillary number, $Ca=\eta_{1} u /\gamma$, which measures the ratio of viscous to capillary forces.  
A suitable unit system is 
$$x^* =  \frac{x}{x_c},\qquad y^*= \frac{y}{x_c} ,\qquad (z^*,h^*)= \left(\frac{z}{h_c},\frac{h}{h_c}\right),$$
\begin{equation}
\langle \vec v  \rangle^* = \frac{\langle \vec v \rangle }{u},\qquad\mathrm{and}\qquad \frac{t}{t_c},
\label{eq:units}
\end{equation}
where $x_c=h_c(3Ca)^{-\frac{1}{3}}$,  
and $t_c=x_c/u.$

\subsection{Parameter values}

The model presented in the previous section is specified by a set of parameters, 
$\rho_1$, $\rho_2$, $\eta_1$, $\eta_2$, $\phi_{\mathrm{eq}}$, $\gamma$, $\xi$ and $M$.  Additionally, 
the fluid wetting properties are specified by the static contact angle, $\theta_e$, which 
depends on $C$.  In the following, we specify the values of these parameters used in our 
simulations (all in lattice units). 

The equilibrium values of the order parameter are fixed to $\phi^* = \pm 1$ by choosing $A=-B$.  
To specify the viscosity of each fluid we fix the mean viscosity $\langle\eta\rangle = 0.1$ and the viscosity contrast, 
$\delta \eta = 0.9$.  The local viscosity, $\eta(\vec r)$, is then set as $\eta_1 = \langle\eta\rangle (1+\delta \eta)$ if $\phi > 0$ 
(viscous film) and $\eta_2 = \langle\eta\rangle (1+\delta \eta)$ if $\phi < 0$ (surrounding low-viscosity 
phase).  The density is set to the same mean value in each phase, $\rho_1 = \rho_2 = 1$.
The fluid-fluid surface tension is fixed to $\gamma = 2.3\times 10^{-3}$.  The interface width is  
set to $\xi = 0.57$, which corresponds to an effective transition zone of $\sim 5$ lattice spacings for the
order parameter profile between the volume values in equilibrium.

In order to ensure that the contact line propagates with a velocity comparable to the velocity of the 
film \cite{Ledesma01}, we fix the mobility to $M=10.0$.  

The equilibrium contact angles used throughout this study correspond to $\theta_e=0^\circ$ for the 
{\it hydrophilic} substrate and $\theta_e=90^\circ$ for the {\it hydrophobic} substrate,  which corresponds to 
$C=0$ and $C=1.7\times 10^{-3}$, respectively. 

The capillary number is set to $Ca=0.41$, which follows from choosing the forcing $f_x$ to give a mean velocity 
value for the film, $u=0.005$.  This value ensures the stability of the LB scheme and a small compressibility 
(density variations of the order of $\sim 1\%$.)

\section{Lattice Boltzmann Method}
\label{sec:LB}
We use the Lattice-Boltzmann (LB) implementation introduced in detail in Ref.~\cite{pagonabarraga,Cates-PhilTransRoySocAMathPhysEng-2005} in the context 
of spinodal decomposition of binary fluid mixtures, and which we have used in a previous work~\cite{Ledesma03}, where we
have focused on the dynamics of unstable thin films in homogeneous solid substrates.  In LB, the dynamics are introduced 
by two sets of velocity distribution functions, ${f_i}$ and ${g_i}$, which evolve in time according to the 
discretized Boltzmann equations,
\begin{equation}
f_i(\vec r + \vec c_i,t + 1) - f_i(\vec r,t) = -\frac{1}{\tau_f}(f_i - f_i^{eq}) + F_i^f,
\label{eq:evf}
\end{equation}
and
\begin{equation}
g_i(\vec r + \vec c_i,t + 1) - g_i(\vec r,t) = -\frac{1}{\tau_g}(g_i - g_i^{eq}),
\label{eq:evg}
\end{equation}
Space is discretized as a cubic lattice where nodes are joined by velocity vectors, $\vec c_i$.  Here we use the D3Q19 model,
which consists of a set of 19 velocity vectors in three dimensions~\cite{pagonabarraga}.  
Hence, $f_i$ and $g_i$ are proportional 
to the number of particles moving in the direction of $\vec c_i$.   
Eqs.~(\ref{eq:evf}) and~(\ref{eq:evg}) are composed of two steps.  First, the distribution functions are relaxed to their equilibrium 
values, represented by $f_i^{eq}$ and $g_i^{eq}$, with relaxation timescales $\tau_f$ and $\tau_g$.  The term $F_i^f$ is related
to the external forcing.  Following this collision stage, distribution functions are propagated to neighboring sites.

The mapping between LB scheme and the hydrodynamic phase-field model is done by defining the hydrodynamic variables 
through moments of the $f_i$ and $g_i$. The local density and order parameter are given by $\sum_if_i=\rho$ and $\sum_i g_i=\phi,$   while 
the fluid momentum and order parameter current are defined as $\sum_if_i \vec c_{i} =\rho \vec v $ and $\sum_ig_i \vec c_{i} =\phi \vec v $.  

Local conservation of mass and momentum is enforced through the conditions $\sum_i f_i = \sum_if^{eq}_i  =\rho $, 
$\sum_i g_i =  \sum_ig^{eq}_i =\phi$, $\sum _i f_i \vec c_i = \sum_if^{eq}_i \vec c_{i} =\rho \vec v $ and $\sum_i g_i \vec c_i = \sum_ig^{eq}_i \vec c_{i} =\phi \vec v $.
In equilibrium, the pressure tensor and chemical potential are defined as $\sum_if^{eq}_i\ \vec c_{i}\vec
c_{i}  =\rho \vec v \vec v + \bar{\bar P}_T$ and $\sum_ig^{eq}_i\ \vec c_i\vec c_i  =\hat M \mu\bar{\bar
\delta} + \phi \vec v \vec v$.

To recover the equilibrium behavior of the phase-field model, the equilibrium distribution functions and the forcing term 
are expanded in powers of $\vec v$\cite{ladd}, \textit{i.e.},
$$f_i^{eq} = \rho \omega_\nu\left(A_\nu^f + 3\vec v\cdot \vec c_{i} + \frac{9}{2}\vec v\vec v : \vec c_{i}\vec c_{i} -\frac{3}{2}v^2 + {\bar{\bar G}}^f:\vec c_{i} \vec c_{i}\right),$$
$$g_i^{eq} = \rho \omega_\nu\left(A_\nu^g + 3\vec v\cdot \vec c_{i} + \frac{9}{2}\vec v\vec v : \vec c_{i}\vec c_{i} -\frac{3}{2}v^2 + {\bar{\bar G}}^g:\vec c_{i} \vec c_{i}\right)$$
and
$$F^f_i = 4\omega_\nu\left(1-\frac{1}{2\tau_f}\right)\left[\vec f \cdot \vec c_i(1+\vec v \cdot \vec c_i)-\vec v \cdot \vec f\right].$$
Here, $\nu$ stands for the three possible magnitudes of the $\vec c_i$ set.  The values of the coefficents are then fixed by the definition 
of the moments of $f_i^{eq}$ and $g_i^{eq}$, and by the symmetry of the lattice. Coefficient values for the D3Q19 LB scheme are
$\omega_0 = 2/9,$ $\omega_1=1/9$ and $\omega_{\sqrt{3}}=1/72;$ $A^f_0=9/2-7/2\mathrm{Tr}\bar{\bar P} ,$
$A^f_1 = A^f_{\sqrt{3}} = 1/\rho \mathrm{Tr}\bar{\bar P}$ and $\bar{\bar G}^f = 9/(2\rho)\bar{\bar P}-
3\bar{\bar \delta}\mathrm{Tr}\bar{\bar P};$  $A_0^g= 9/2-21/2\hat M\mu,$  $A_1^g = A^g_{\sqrt{3}} = 3\hat M\mu/\rho$ and
$\bar{\bar G}^g = 9/(2\rho)\hat M\mu(\bar{\bar1}-\bar{\bar\delta}),$ where $\bar{\bar 1}$ is the unit matrix.

The hydrodynamic phase-field model, given by Eqs.~(\ref{eq:Continuity}), ~(\ref{eq:NavierStokes}) and~(\ref{eq:ConvectionDiffusion}), 
up to second order corrections in the velocity, can be recovered 
by performing a Chapman-Enskog expansion of Eqs.~(\ref{eq:evf}) and~(\ref{eq:evg})\cite{ladd}.  The LB 
scheme maps to the hydrodynamic model through the relaxation timescales, \textit{i.e.}, $\eta=(2\tau_f - 1)/6$
and  $M=(\tau_g - 1/2)\hat M,$ and through the body force $\vec f = \rho \vec g$.

Solid boundaries in the Lattice-Boltzmann method are implemented by means of the well known bounce-back
rules\cite{ladd,Pagonabarraga02}.  In the lattice nodes that touch the solid, the propagation scheme is
modified so the distribution functions are reflected to the fluid rather than absorbed by the solid.  As a
consequence, a stick condition for the velocity is recovered approximately halfway from the fluid node to the
solid node.

\section{Results}

\label{sec:Results}

\begin{figure}
\begin{centering}
\includegraphics[width=0.45\textwidth]{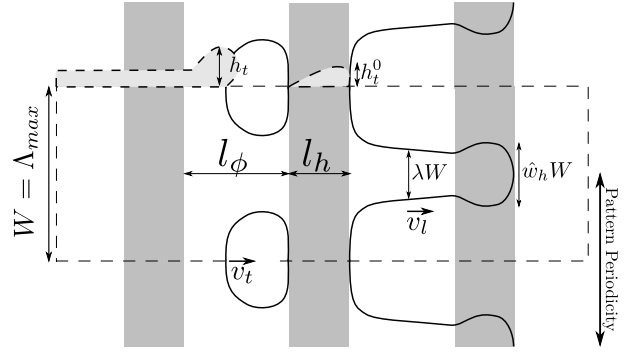}
\end{centering}
\caption{Geometry of the transverse stripe pattern.  The pattern is composed of alternated hydrophilic (grey) and hydrophobic (white) stripes 
of lengths $l_h$ and $l_\phi$, respectively.  The panel shows a schematic representation of the contact line (solid line), which spreads 
on the hydrophilic stripes up to a width $\hat w_h W$ before advancing on the next hydrophobic stripe.  Spreading causes that a series of dry spots
are left on the hydrophobic stripes, which are eventually covered by the trailing edge of the film.  The dotted lines depict the thin film profile, which 
has a typical thickness $h_t^0$ on a wet hydrophilic stripe, and a maximum thickness $h_t$ at the trailing edge on a hydrophobic stripe.   At the grey 
zones $\theta_e=0^\circ$ while at the white zones $\theta_e=90^\circ$.  \label{fig:tspattern}}
\end{figure}

\begin{figure}
\begin{centering}
\includegraphics[width=0.45\textwidth]{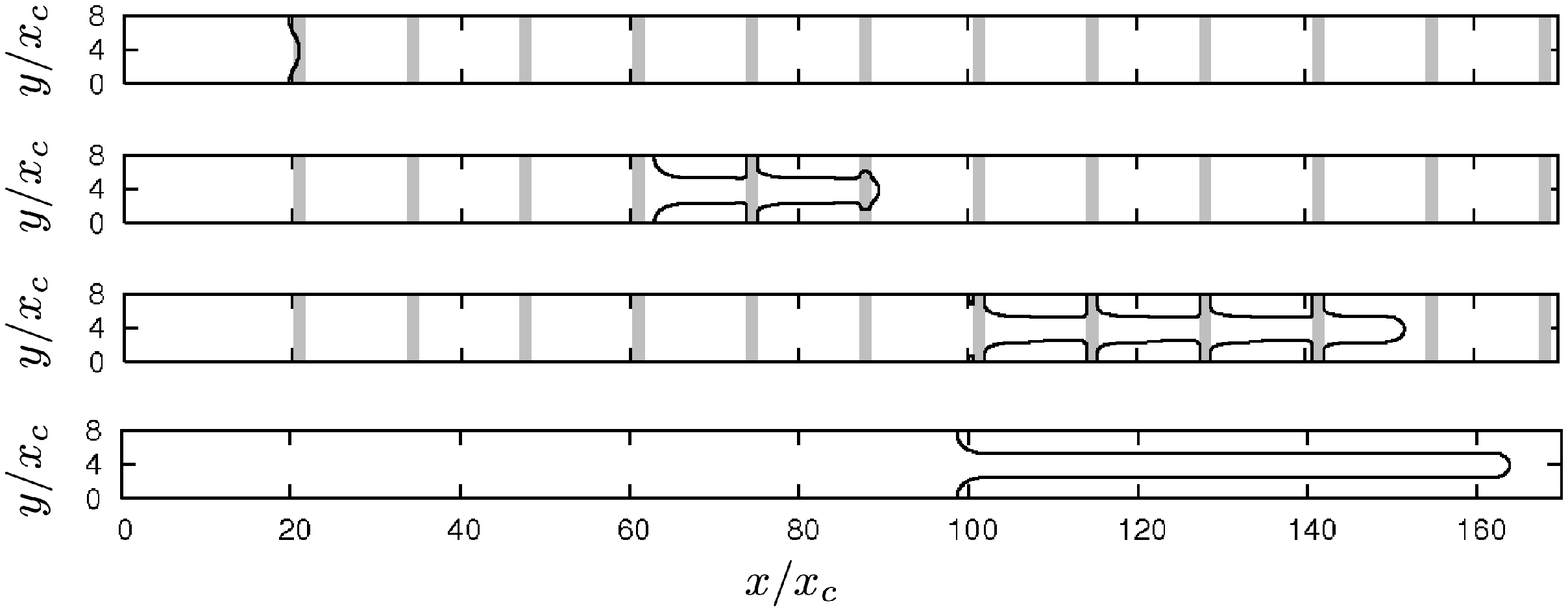}\\

(a)\\

\includegraphics[width=0.45\textwidth]{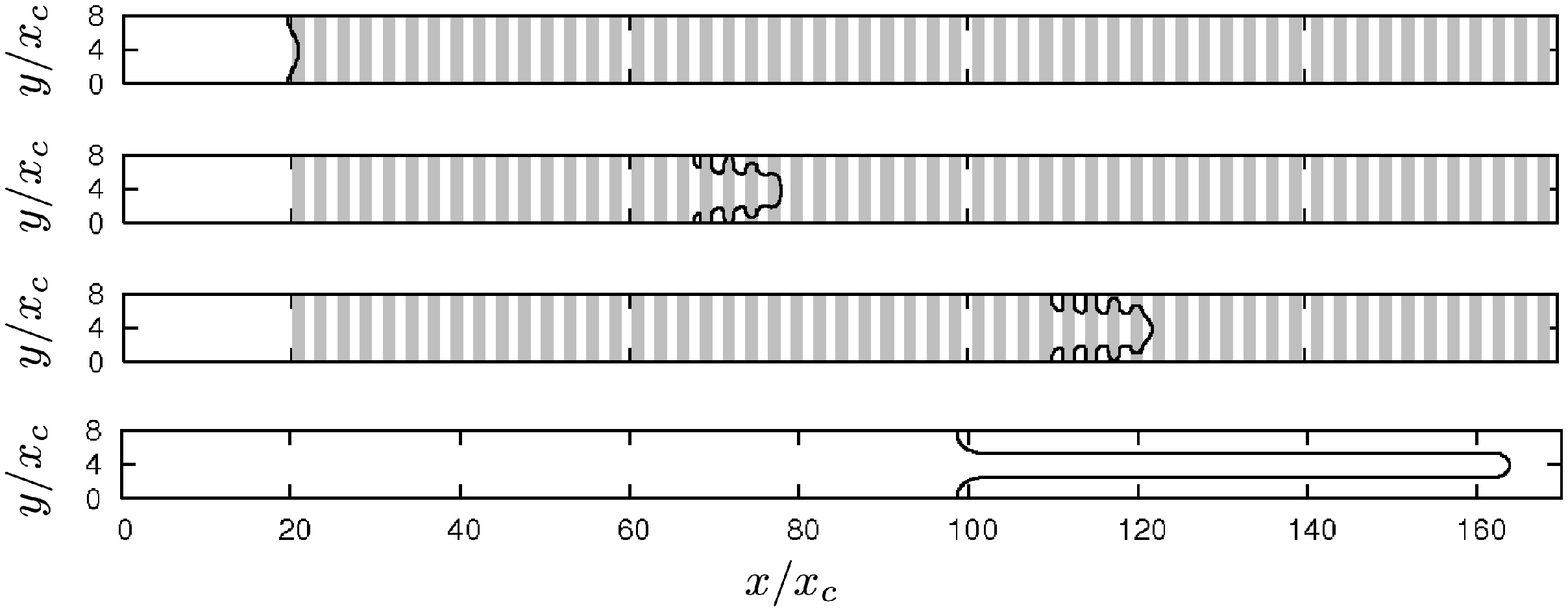}\\

(b)\\

\end{centering}
\caption{Contact line evolution on transverse stripe patterns for hydrophilic stripes of length $l_h/x_c=1.3$ at
different covering fraction, $f$.  The first panel in each figure shows the initial perturbation at $t/t_c=0$, while the subsequent 
two panels show the contact line at time intervals of $\Delta t/t_c = 67$.  For comparison, the last panel in (a) and (b) shows the finger that grows 
on a homogeneous hydrophobic substrate ($f$=0) at $t= 2\Delta t /t_c$. 
For both values of $f$, the contact line grows as a finger on the hydrophobic stripes, and spreads transversely 
on the hydrophilic domains. (a) For a small fraction of hydrophilic 
stripes, $f=0.11$, the contact line grows at a smaller rate than the one measured on a homogeneous 
substrate.  (b) Increasing the fraction to $f=0.55$ has the effect of decreasing the growth 
rate dramatically, making the length of the contact line substantially smaller than the length of the unperturbed finger.  
As seen by comparing the position of the contact line in third panel of (a) and (b), the reduction of the growth rate arises both from a slow down of the leading 
edge and a speed up of the trailing edge of the contact line. \label{fig:tstripes}}
\end{figure}

We want to assess if a small fraction of hydrophilic domains can induce a significant reduction of the 
growth of the contact line.  We thus characterize the transverse stripe arrangement by the width of the hydrophilic stripes, 
$l_h$, and the fraction of the substrate that is covered by them,  $f=l_h/(l_h+l_\phi)$, where $l_\phi$ is the length of the 
hydrophobic stripes, as depicted in Fig.~\ref{fig:tspattern}.  

In a previous study \cite{Ledesma03}, we have considered the motion of thin films on homogeneous substrates.  We have found that 
the front develops as a finger if the substrate is hydrophobic ($\theta_e=90^\circ$), with an inter-finger 
spacing given by $\Lambda_{max} (\theta_e=90^\circ)\simeq 8 x_c$.  In contrast, for a hydrophilic substrate the front 
develops as a sawtooth with a typical lateral lengthscale of $\Lambda_{max}(\theta_e=0^\circ) \simeq 14 x_c$.   Due to the 
periodicity of the emerging structures, it is sufficient to consider simulation domains of width $W = \Lambda_{max} (\theta_e=90^\circ)\simeq 8 x_c $. 
Larger system sizes obeying $W = n \Lambda_{max} (\theta_e=90^\circ) $, with $n$ being an integer, only give rise to additional 
identical fingers, whose effect is already taken into account by the periodic boundary conditions in our simulations.  For non-integer
$n$ we expect a non-linear competition of emerging fingers, which we have observed previously \cite{Ledesma03}.  However, this competition 
should not modify significantly the main effect of the transverse-stripe pattern.  

In the following, we will explore the dynamics of the front when varying $l_h$ and $f$.  We study the effect 
of the occupation fraction in the range $0 \leq f \leq 1$, while for the length of the stripes we choose the values
$l_h/x_c=1.3$, $l_h/x_c=4.0$, and $l_h/x_c=6.7$.    As before, we consider the evolution of a single 
mode in systems whose width is chosen to be close to the most unstable wavelength predicted for the hydrophobic substrate.

Figures~\ref{fig:tstripes}(a) and~\ref{fig:tstripes}(b) show the evolution of the contact line for $f=0.11$ and $f=0.55$ 
respectively, for a stripe width $l_h/x_c=1.3$ at three subsequent times, $t_0/t_c=0$, $t_1/t_c=67$ and $t_2/t_c=134$.
For comparison, the last panel at the bottom of (a) and (b) shows the interface profile for a homogeneous hydrophobic 
substrate ($f$=0) at $t_2/t_c=134$.  The main feature in these patterns is that the growth rate of the contact line decreases strongly 
as the fraction of hydrophilic 
stripes is increased.   For the value of $l_h$ considered in Fig.~\ref{fig:tstripes}, the growth rate actually vanishes at a 
critical fraction $f_c\simeq 0.59$.   We first notice that for both patterns shown in Fig.~\ref{fig:tstripes}, the leading 
edge of the contact line grows as a finger 
when it is in contact with a hydrophobic stripe. The finger grows until it touches a hydrophilic stripe, where it starts 
to spread transversely.  As spreading takes place, the leading edge continues advancing over the hydrophilic domain, until it reaches the next hydrophobic
stripe, where it develops as a finger again.   A closer inspection of Fig.~\ref{fig:tstripes}(a) shows that a series of spots are left 
uncovered momentarily on the hydrophobic stripes after the leading edge has passed by. This is a consequence of the 
transverse spreading process that occurs every time the leading edge passes over a hydrophilic stripe.  The second and third panels in  
Fig.~\ref{fig:tstripes}(a) show that the trailing spot is eventually covered by the film, and that only after 
the trailing spot has disappeared, the following spot starts to be covered.  This mechanism holds 
for larger $f$, as can bee seen in Fig.~\ref{fig:tstripes}(b).    The global result is that the transverse pattern 
induces a slow down of the leading edge, as can be seen by comparing the 
contact line with the finger that grows on a homogeneous substrate.  This effect is stronger for $f=0.55$ than for $f=0.11$.  
Additionally, the transverse pattern induces a speed up of the trailing edge,  and effect that can also clearly be seen in Fig.~\ref{fig:tstripes}.
Again, this effect is stronger the larger $f$ is.

The combination of the speed up of the trailing edge and the slow down of the leading edge causes the contact line growth rate, 
$\dot L$, to decrease with increasing fraction of hydrophilic stripes.  Figure~\ref{fig:globratefw} shows a plot of $\dot L$ as a function 
of $f$.  The limits in this plot correspond to the finger ($f=0$) and sawtooth ($f=1$) solutions on homogeneous substrates.   For small $f$, 
the growth rate converges to the growth rate of the unperturbed finger, while increasing $f$ has the effect of
decreasing the growth rate as explained above.  We find that $\dot L$ vanishes at a critical 
fraction, $f_c(l_h)$, which is considerably smaller than unity.  This means that the transverse pattern 
induces the saturation of the contact line by caused by transverse spreading of the film on hydrophilic domains. 

To analyze the effect of the length of the stripes, $l_h$, we have carried out additional simulations at fixed $f$ while varying 
$l_h$ and have measured the corresponding growth rate.  We plot these results in Fig.~\ref{fig:globratefw}, where we 
observe that the growth rate decreases systematically as $l_h$ is increased.  Accordingly, the critical fraction $f_c(l_h)$
is a decreasing function of $l_h$.

\subsubsection*{Analytical model for growth on transverse stripe patterns}
\begin{figure}
\begin{centering}
\includegraphics[width=0.45\textwidth]{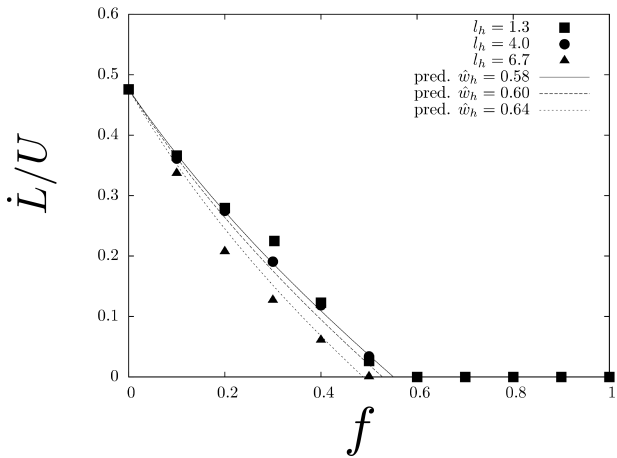}\\
\end{centering}
\caption{Global contact line growth rate as a function of the fraction of hydrophilic stripes for 
different hydrophilic stripe lengths on transverse striped patterns.  The growth rate $\dot L$, decreases 
as the fraction of hydrophilic stripes, $f$, increases.   For fixed $f$, increasing the length of the hydrophilic stripes,  $l_h$,
has the effect of decreasing the growth rate.   Continuous lines correspond to the kinematical theoretical prediction, 
for which the dependence on $l_h$ is contained in the parameter $\hat w_h$, which we measure from simulations.  The plot
shows that the growth rate vanishes at a fraction that is smaller than unity. 
 \label{fig:globratefw} }
\end{figure}

In order to gain insight into the growth of the contact line on transverse patterns, we will analyze the interplay 
between hydrophilic and hydrophobic stripes using a kinematical argument.  

Let us denote $\Delta t_l$ and $\Delta t_t$ the time intervals in which the leading and trailing edges 
sweep a distance $\Delta l = l_\phi+l_h$, respectively.  For a mean steady growth of the interface, each of these intervals 
can be decomposed as the sum of the time that the interface spends on a hydrophilic and a hydrophobic stripe.  
We thus write
\begin{equation}
\Delta t_l = \Delta t_l^h + \Delta t_l^\phi \qquad  \mathrm{and} \qquad \Delta t_t = \Delta t_t^h + \Delta t_t^\phi,
\label{eq:Times}
\end{equation}
where superscripts $h$ and $\phi$ correspond to the hydrophilic and hydrophobic domains, respectively. 

For sufficiently long hydrophobic stripes, the thin film should grow as an unperturbed finger, with 
leading and trailing edge velocities $v_l$ and $v_t$, respectively.  Disregarding the transient relaxation 
to these values, it is reasonable to assume that the contact line changes its velocity instantaneously 
to $v_l$ and $v_t$ as soon as it comes into contact with a hydrophobic stripe.  From this assumption, it follows 
that the residence time of the leading and trailing edges of the film in a hydrophobic 
stripe are 
$$\Delta t_l^\phi=\frac{l_\phi}{v_l} \qquad  \mathrm{and} \qquad \Delta t_t^\phi=\frac{l_\phi}{v_t}.$$

For the hydrophilic stripes, we can write similar expressions for the residence times of the leading 
and trailing edges of the interface in a hydrophilic stripe, 
$$\Delta t_l^h=\frac{l_h}{v_l^h} \qquad  \mathrm{and} \qquad \Delta t_t^h=\frac{l_h}{v_t^h}.$$
The velocity $v_l^h$ corresponds to the mean velocity of the leading edge on a hydrophilic stripe,  
while $v_t^h$ is defined as the velocity of the jump that the trailing 
interface undergoes on a prewet hydrophilic stripe.      

To estimate $v_l^h$, we first approximate the volumetric flux supplied by the finger to the hydrophilic stripe 
as $\dot q_l \sim h_f  \lambda W v_l, $ where $h_f$ and $\lambda W$ are the thickness and width of the finger 
in the hydrophobic stripe, respectively (see Fig.~\ref{fig:tspattern}).   
The flow rate $\dot q_l$ sustains the spreading of the leading edge, which covers a volume $V_l \simeq h_f l_h  \hat w_h W$ 
of the hydrophilic stripe before it moves over the next hydrophobic stripe.  
The volume $V_l$ depends on $\hat w_h$, which is the lateral fraction of the stripe that the film covers before the 
leading edge advances over the next hydrophobic domain. In writing the expression for $V_l$ we have assumed that the film 
thickness does not vary significantly during spreading. The residence time of the spreading process is therefore $\Delta {t_l}^h = V_l / \dot q_l ,$
which gives a mean spreading velocity  
\begin{equation}
{v_l}^h \simeq \frac{l_h}{\Delta{t_l}^h} = \frac{\lambda}{\hat w_h} v_l.
\label{eq:vlh}
\end{equation}

We next estimate the jumping velocity of the trailing edge, $v_t^h$.  This velocity 
corresponds to the ratio between the length of the hydrophilic stripe, $l_h$, and
the waiting time to observe the jump in the contact line position, $\Delta{t_t}^h$.  Let $h_t$ be the
thickness of the trailing edge of the film when it touches the hydrophilic stripe, which is already wet and has a local thickness $h={h_t}^0$, 
as shown in Fig.~\ref{fig:tspattern}.  After contact, 
the cross section of the film in the $x-z$ plane will evolve as ${A_t}^h\simeq h l_h={h_t}^0l_h + \dot q_t t,$ 
where the flow rate per unit length in the $y$ direction, $\dot q_t$, obeys $\dot q_t \sim v_t h_t$.  We expect that ${A_t}^h$ grows until the thickness of the film is $h=h_t$, from which it follows that the corresponding waiting time 
is 
\begin{equation}\Delta {t_t}^h = \frac{(h_t -{h_t}^0)l_h}{v_th_t}.
\end{equation}
We thus obtain an estimate for the jumping velocity of the trailing edge,
\begin{equation}
\hat v_t^h = \frac{v_t}{1-\hat h_t},
\label{eq:vth}
\end{equation}
where $\hat h_t={h_t}^0/h_t.$

We now return to Eq.~(\ref{eq:Times}), which we write in terms of the mean velocities of the leading 
and trailing edges, $\hat v_l $ and $\hat v_t$, {\it i.e.},
\begin{equation}
\frac{l_\phi+l_h}{\hat v_l}= \frac{l_h}{{v_l}^h}+\frac{l_\phi}{v_l}\qquad \mathrm{and} \qquad \frac{l_\phi+l_h}{\hat v_t} = \frac{l_h}{{v_t}^h}+\frac{l_\phi}{v_t}.
\label{eq:TSvels}
\end{equation}
From Eq.~(\ref{eq:TSvels}) we obtain the mean velocities of the interface as a function of the 
local velocities and the fraction of hydrophilic stripes, 
\begin{equation}\hat v_l = \frac{v_l}{1+f\left(\frac{v_l}{{v_l}^h}-1\right)} \qquad \mathrm{and} \qquad
\hat v_t = \frac{v_t}{1+f\left(\frac{v_t}{{v_t}^h}-1\right)},
\end{equation}
where $v_l$ and $v_t$ can be measured from the single finger solution, while $v_l^h$ and $v_t^h$ 
follow from Eqs.~(\ref{eq:vlh}) and~(\ref{eq:vth}).  For the leading edge velocity we obtain
\begin{equation}
\hat v_l = \frac{v_l}{1+f\left(\frac{\hat w_h}{\lambda }-1\right)},
\label{eq:vltrans}
\end{equation}
which is proportional to the leading edge velocity of the finger and depends on the fraction of
hydrophilic stripes, as well as on the degree of transverse spreading, which is contained in the
dependence on $\hat w_h$. 

For the trailing edge velocity we find
\begin{equation}
\hat v_t = \frac{v_t}{1-\hat h_tf},
\label{eq:vttrans}
\end{equation}
which is controlled by the local thickness of the film on the prewet hydrophilic stripe. 

Equations~(\ref{eq:vltrans}) and~(\ref{eq:vttrans}) can be used to obtain the growth rate of the 
contact line,
\begin{equation}
\dot L = \hat v_l - \hat v_t = \frac{v_l}{1+f\left(\frac{\hat w_h}{\lambda}-1\right)} -  \frac{v_t}{1-\hat h_t f}.
\label{eq:ldottrans}
\end{equation}
In order to verify the validity of Eq.~(\ref{eq:ldottrans}), values of $\lambda$, $v_f$, $v_t$, $\hat w_h$, and $\hat h_t$
have to be provided.   The finger width, $\lambda$, and the finger velocities of the leading and trailing edges, $v_f$, and $v_t$,
are known from the finger growth on a homogeneous hydrophobic substrate.  For our simulation parameters, $v_l=1.1$, $v_t=0.6$, and 
$\lambda \simeq 0.3$ as measured from the finger solution ($f=0$).    On the other hand, $\hat h_t$ and $\hat w_h$ are
geometrical parameters that can be measured from simulations.  We find that $\hat h_t$ does not depend on the length 
or on the fraction of hydrophilic domains and we measure $\hat h_t = 0.33$.   In contrast,  $\hat w_h$ shows 
a dependence on $l_h$, which is essentially controlled by the competition
between the forward motion of the film the lateral spreading caused by the hydrophobic covering.
For long hydrophilic domains, the film
spreads to a larger extent before the leading edge arrives to the next hydrophobic domain.  
Therefore, $\hat w_h$ increases with $l_h$.  This introduces the dependence of the growth rate 
on $l_h$.    We do not have a prediction for $\hat w_h$ as a function of $l_h$.  Instead, we measure 
these values from our simulations.  
For the runs performed we find $\hat w_h = 0.58$ for $l_h/x_c=1.3$, $\hat w_h =0.60$  for $l_h/x_c=4.0$, and 
$\hat w_h =0.64$ for $l_h/x_c=6.7$.  

In Figure~\ref{fig:globratefw} we show a comparison of Eq.~(\ref{eq:ldottrans}) with simulation results.   As depicted in the 
figure, the theoretical prediction captures the main 
dependence of the growth rate of the contact line, $\dot L$, both on $f$ and, through $\hat w_h$, on $l_h$.  For a fixed hydrophilic 
domain length, $l_h$, $\dot L$ decreases with increasing $f$ due both to a decrease of the leading edge velocity, as shown by 
Eq.~(\ref{eq:vltrans}), and by an increase of the trailing edge velocity, 
as predicted by Eq.~(\ref{eq:vttrans}).  On the other hand, 
for fixed  $f$, increasing $l_h$ has the effect of decreasing the growth rate.   This effect can be traced back to Eq.~(\ref{eq:vlh}), 
which shows that increasing $\hat w_h$ has the effect of slowing down the leading edge on a hydrophilic domain.   
\begin{figure}
\begin{centering}
\includegraphics[width=0.45\textwidth]{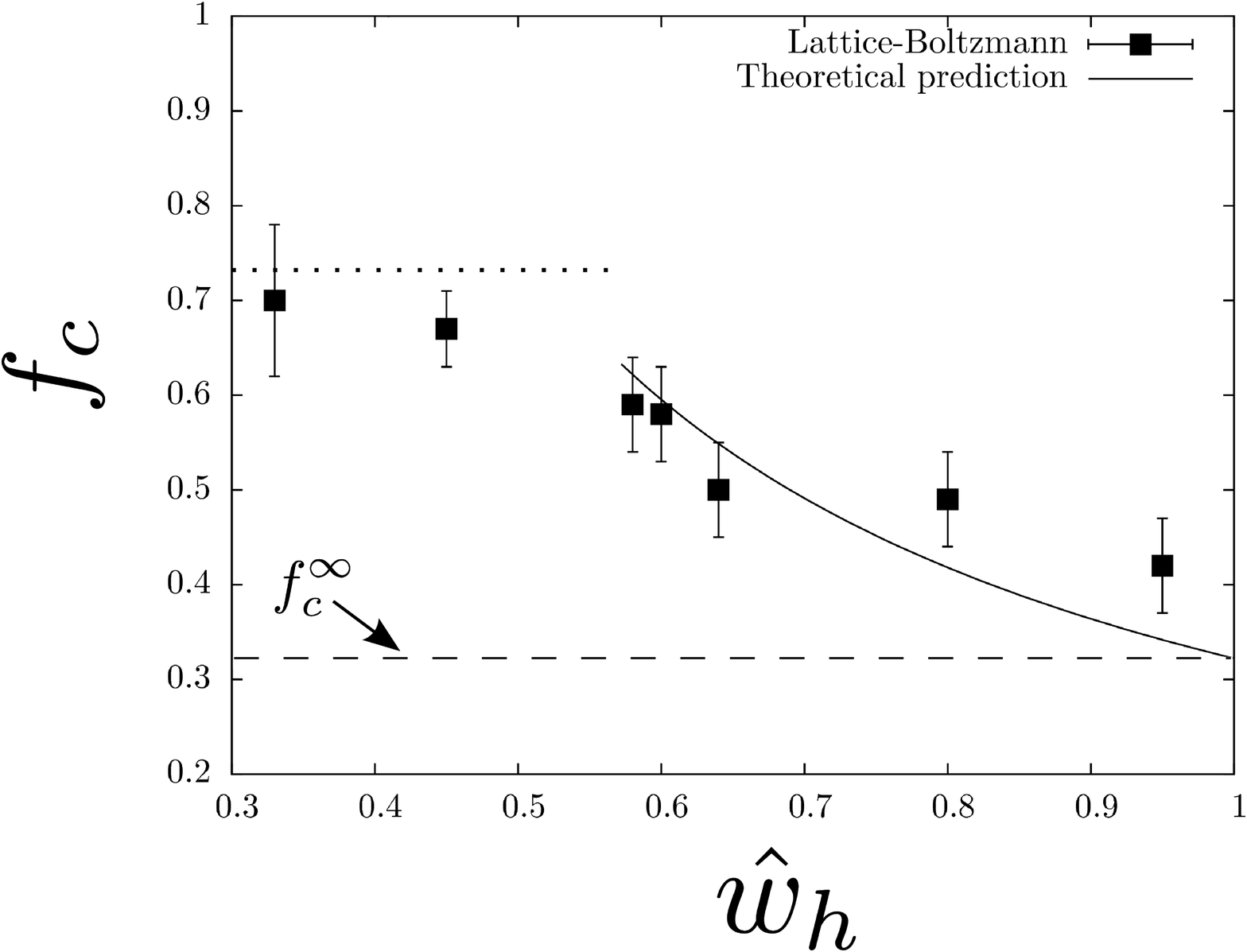}\\
\end{centering}
\caption{Critical fraction, $f_c$, as a function of 
the fraction of the hydrophilic stripe covered by the leading edge, $\hat w_h$.  The fraction $\hat w_h$ increases with the length of the hydrophilic stripes, $l_h$.  
Lattice-Boltzmann simulations (solid symbols) show that the critical fraction decreases substantially with increasing $\hat w_h$, {\it i.e.}, with increasing $l_h$.  This behavior is captured by a kinematical model for large $\hat w_h$, depicted by the solid line (see text).  
 \label{fig:criticalfraction} }
\end{figure}

The theoretical result given by Eq.~(\ref{eq:ldottrans}) predicts a critical fraction 
of hydrophilic stripes, $f_c <  1$, above which the growth rate, $\dot L$, vanishes. 
Such a saturation emerges both from  the decrease of the leading edge velocity and the 
increase of the trailing edge velocity, an effect that is intimately linked to the fraction of 
hydrophilic stripes imposed to the 
substrate.    The critical fraction follows from setting $\dot L = 0$ in Eq.~(\ref{eq:ldottrans}), and reads
\begin{equation}
\label{eq:fc}
f_c=\left(v_l-v_t\right)\left[v_l\hat h_t+v_t\left(\frac{\hat w_h}{\lambda}-1\right)\right]^{-1}.
\end{equation}
Here, the relevant dependence is $f_c(\hat w_h)$, given that $\hat w_h$ can be controlled by choosing the length 
of the hydrophilic domains $l_h$.

In Fig.~\ref{fig:criticalfraction} we show a plot of $f_c$ as a function of $l_h$ and $\hat w_h$.  We have included   
$f_c$ data for four additional values of $l_h$; $l_h/x_c=0.4$, $l_h/x_c=0.7$, $l_h/x_c=10.1$ and $l_h=13.4$.   
Corresponding values of $\hat w_h$ are $\hat w_h=0.33$, $\hat w_h=0.45$, $\hat w_h=0.80$ and $\hat w_h=0.95$, respectively.    
We also show the theoretical prediction given by Eq.~(\ref{eq:fc}), which accurately captures the decrease 
of $f_c$ as $\hat w_h$ increases for large $\hat w_h$.   

An interesting limit corresponds to short hydrophilic domains, which we can include in our theory. 
As shown in the figure, the critical fraction reaches a saturation value,  which originates from a weaker lateral 
spreading.   In this limit the lateral size of the film in a hydrophilic 
domain remains close to the width of the finger, {\it i.e.,} $\hat w_h \simeq \lambda$. Still, the hydrophilic 
interaction causes a decrease in the local contact angle.  As the leading edge encounters the next hydrophobic boundary, the
capillary ridge grows, until the contact angle reaches the advancing contact angle of the hydrophobic
stripe.  The relevant timescale is given by the filling of the capillary ridge, which we approximate as a spherical 
cap of radius $R \simeq h_f/2$ and 
volume $V_l=2\pi(h_f/2)^3/3.$  The volumetric flow coming into the sphere still obeys $\dot q_l=h_f\lambda W v_l.$
Adding this contribution to the residence time, we find the spreading-filling velocity, 
\begin{equation}
v_l^h (\hat w_h \rightarrow \lambda) = \frac{v_l}{\frac{\pi h_f^2}{6\pi \lambda W l_h} + \frac{\hat w_h}{\lambda}}.
\label{eq:vlhlim}
\end{equation}

Eq.~(\ref{eq:vlhlim}) can be used to obtain the critical fraction in this limit, which reads
\begin{equation}
\label{eq:fclim}
f_c (\hat w_h \rightarrow \lambda ) = \left(v_l-v_t\right)\left[v_l\hat h_t+v_t\left(\frac{\pi h_f^2}{6\pi \lambda W l_h^*}\right)\right]^{-1},
\end{equation}
where $l_h^*$ is the crossover hydrophilic stripe length below which 
filling dominates over spreading.  The dotted line in Fig.~\ref{fig:criticalfraction} 
shows the value predicted by Eq.~(\ref{eq:fclim}) for $l_h^*/x_c\approx0.4$, which is a typical stripe length 
for which no lateral spreading is observed.  As depicted in the figure, this limiting value 
reasonably captures the saturation of the critical fraction for small $\hat w_h$.  It is important 
to remark that the estimate of the leading edge velocity given by Eq.~(\ref{eq:vlhlim}) is 
expected to be valid only as $\hat w_h \rightarrow \lambda$; for larger $\hat w_h$, the shape of the leading 
ridge is expected to depend on $\hat w_h$.  Still, the filling timescale is subdominant, as shown by the good agreement
of simulation data with Eq.~(\ref{eq:fc}), and we thus disregard it. 

The opposite case, corresponding to  $\hat w_h\rightarrow 1,$ is an interesting regime, 
as in this case the imposed pattern has the smallest critical fraction of hydrophilic stripes for which saturation is 
expected.  In this limit the leading edge spreads completely in the transverse direction 
on a hydrophilic stripe, a situation favored by long hydrophilic domains.  In Eq.~(\ref{eq:fc}), this 
is equivalent to to setting $\hat w_h=1$,  from which it follows that the minimum possible critical fraction is 
\begin{equation}
f_c^{\infty	} =\left(v_l-v_t\right)\left[v_l\hat h_t+\left(\frac{1}{\lambda}-1\right)v_t\right]^{-1}.
\end{equation}
For the simulation values used in this work we find $f_c^{\infty}\simeq 0.3$.  In Fig.~\ref{fig:criticalfraction}(b) we plot 
this value, which is a good estimate of the limiting critical fraction as suggested by Lattice-Boltzmann results.  
While both simulation and theory predict a slow final decay towards the limiting critical fraction, there is a rapid variation
for intermediate stripe lengths, as shown in Fig.~\ref{fig:criticalfraction}.  For example, for $l_h/x_c=13.4$ we obtain 
$f_c \simeq 0.4$. 
This value of the stripe length should be compared to the the typical finger spacing, which for our simulations 
is $\Lambda_{max}/x_c=8$. 
Therefore, one expects critical fractions significantly smaller than unity for stripes whose length is comparable to the 
lengthscale of the unstable film.

\section{Conclusions}

\label{sec:DC}
By means of Lattice-Boltzmann simulations and analytic approximations, we have studied the effect of 
heterogeneous hydrophilic-hydrophobic substrates composed of transverse striped patterns on the 
dynamics of forced thin liquid films, and have demonstrated that the film growth can be controlled by 
the interaction of the film with a prescribed hydrophilic-hydrophobic pattern. 

We have focused on a scenario where the unstable contact line gives rise to 
sawtooth and finger structures on homogeneous hydrophilic and hydrophobic substrates,
respectively, and have examined how the growth of the contact line  is altered by the 
chemical pattern.    To this end, we have considered patterns where the typical lengthscale 
is comparable to the most unstable wavelength of the contact line, $\Lambda_{max}$.  

For the transverse striped patterns considered, we have shown that above a critical fraction 
of hydrophilic stripes, $f_c$, the saturation of the film is achieved.  
The mechanism that causes such a saturation originates from transverse spreading 
of the leading edge of the contact line on the hydrophilic domains and from the growth 
of the film at the hydrophilic-hydrophobic boundary so as to reach the advancing contact angle.  
These processes contribute to a reduction of the velocity of the leading edge.  As this occurs, a film of fluid 
is left on the hydrophilic domains, thus increasing the velocity of the trailing edge as soon as it comes into contact 
with a hydrophilic stripe, which already has been wet. The trailing edge jumps every time it finds a wet domain, with 
the corresponding gain in the trailing edge velocity.  The global outcome is that for sufficiently large fractions of 
hydrophilic stripes, the net growth rate of contact line vanishes, independently of the details of the dynamics 
of the contact line.

We have captured this effect with a kinematical model, assuming local relaxation of the leading and 
trailing edge velocities to `hydrophilic' and `hydrophobic' values.  Hydrophobic values have been approximated 
as the velocities of the leading and trailing edges of an unperturbed finger, while the hydrophilic velocity values 
have been estimated by mass conservation.  The global growth rate of the contact line, $\dot L$,  has been 
determined by the difference of the mean velocities of the leading and trailing edges of the contact line.  
By estimating these velocities on kinematical grounds, we have obtained a prediction for the global growth 
rate, which is given by Eq.~(\ref{eq:ldottrans}).   We have estimated the critical fraction, $f_c$, 
at which saturation occurs indirectly as a function of the length of the hydrophilic domains, given by Eq.~(\ref{eq:fc}).  
For long hydrophilic domains, in which transverse spreading is more efficient,  we find a good agreement between the simulation 
results and the theoretical prediction.  In this limit, we have estimated the minimum fraction of hydrophilic stripes for which 
saturation is expected, which for the simulation parameters considered in this work is as low as $f_c\simeq 0.3$.  
However, experimentally feasible domain lengths, comparable to the length scale of the growing fingers, give critical 
fractions of about $f_c=0.6$ to $f_c=0.7$, still considerably smaller than unity.  

Our theoretical prediction, given by Eq.~(\ref{eq:fc}) can be used to estimate the critical fraction of 
hydrophilic stripes above which the saturation of the contact line is expected.  
To examine the reliability of our approach in a real system we consider a potential microfluidic
device where a water film of thickness $h_c=10~\mu\mathrm{m}$ is forced on a micropattern in the presence
of air. The air-water pair closely reproduces the viscosity contrast used in the simulations. 
Our simulations show that the instability is effectively suppressed using a sharp contrast in the wetting
properties between the domains forming the micropattern.  For example, a front moving over a Polymethyl 
methacrylate (PMMA) substrate would have a relatively large contact angle ($\theta_e \approx 74^\circ$), where fingers
are expected to develop \cite{Ledesma03}.
In that case, the low-wetting angle domains could be formed using glass ($\theta_e \approx 25^\circ$)
or a gold coating ($\theta_e \approx 0^\circ$) \cite{Bruus01}.

Microfluidic devices operate over a wide range of velocities, covering from $\mu \mathrm{m/s}$ to 
$\mathrm{cm/s}$ \cite{Squires-RevModPhys-2005}.
Here we consider an injection velocity of $u=1~\mathrm{mm/s}$ as a representative example.  Under these
circumstances, the required body force to drive the film follows from the force balance far
from the contact line, $f_x = 3\eta_1 u /h_c^2 \sim 10^{4}~\mathrm{Pa/m}$.  Such body forces are of the order
of the gravitational force density for water and can be accomplished by simple column devices and accurately controlled by
syringe pumps.  From our simulation measurements we estimate the leading and trailing edge velocities of
steady finger propagating at the imposed speed on a hydrophobic substrate, $v_l\approx  1.1 ~ \mathrm{mm/s}$ and
$v_l\approx 0.6 ~ \mathrm{mm/s}$.  Our prediction shows that the film will saturate at a different fraction of hydrophilic
coverage depending on the width of the hydrophilic stripes, $l_h$, which determines the fraction of lateral spreading, $\hat w_h$.
A sensible choice is to set $l_h$ to the order of magnitude of the lengthscale of the instability,
$x_c = h_c(3Ca)^{-1/3}$.
Thus, for $l_h = x_c \sim 10^2~\mu \mathrm m$, we get the spreading fraction $\hat w_h \approx 0.58$.  Using this value of $w_h$ and
comparing with Fig. 4, the film is expected to saturate at a critical fraction $f_c \approx 0.6$.

The previous example shows the experimental feasibility of the mechanisms proposed in this work for a regular liquid under typical
conditions found in microfluidics.
However, the fact that the relevant lengths scale with the film thickness (expressed here by the choice of units to report our results) indicates
that a proper up- or down-scaling of the system size and the applied forcing can provide a substantial flexibility to induce analogous phenomena 
in different liquid/gas or liquid/liquid mixtures
at scales below the capillary length.  This opens the possibility of a better control of the fingering instability in controlled environments 
by substrate heterogeneity.
Our approach relies on chemical patterning, which has already been explored experimentally, even at the microscale. 
We therefore expect our results to motivate further experimental investigations, with potential impact in microfluidic
technologies. 

\label{sec:Conclusions}

\section{Acknowledgments}
R.L-A. wishes to thank M. Pradas for useful discussions.  We acknowledge financial support from Direcci\'on General de Investigaci\'on (Spain) under projects FIS\ 2009-12964-C05-02, FIS\ 2008-04386, and DURSI projects SGR2009-00014 and SGR2009-634.   R.L.-A. acknowledges support from CONACyT (M\'exico) and Fundaci\'on Carolina(Spain).    The computational work presented herein has been carried out in the MareNostrum Supercomputer at Barcelona Supercomputing Center.  \label{sec:Acknowl}

\end{document}